\documentclass[12pt]{article}

\usepackage{amsmath}
\usepackage{amssymb}
\usepackage{amsfonts}
\usepackage{latexsym}
\usepackage{color}

\catcode `\@=11 \@addtoreset{equation}{section}

\catcode `\@=12

\newcommand{\be}{\begin{equation}}
\newcommand{\en}{\end{equation}}
\newcommand{\bea}{\begin{eqnarray}}
\newcommand{\ena}{\end{eqnarray}}
\newcommand{\beano}{\begin{eqnarray*}}
\newcommand{\enano}{\end{eqnarray*}}
\newcommand{\bee}{\begin{enumerate}}
\newcommand{\ene}{\end{enumerate}}

\newcommand{\mc}{\mathcal}

\newcommand{\Sc}{{\cal S}}
\newcommand{\E}{{\cal E}}
\newcommand{\F}{{\cal F}}

\newcommand{\1}{1 \!\! 1}

\newcommand{\Hil}{\mc H}

\catcode `\@=11 \@addtoreset{equation}{section}
\catcode `\@=12

\begin{document}

\thispagestyle{empty}

\vspace*{2cm}

\begin{center}
{\Large \bf  Matrix computations for the dynamics of fermionic systems}\\[10mm]

{\large F. Bagarello}\\
  DEIM,
Facolt\`a di Ingegneria,\\ Universit\`a di Palermo, I-90128  Palermo, Italy\\
e-mail: fabio.bagarello@unipa.it\\ Home page:
www.unipa.it/fabio.bagarello\\

\vspace{3mm}

\end{center}

\vspace*{2cm}

\begin{abstract}
\noindent In a series of recent papers we have shown how the dynamical behavior of certain classical systems can be analyzed using operators
evolving according to Heisenberg-like equations of motions. In particular, we have shown that raising and lowering operators play a relevant role in this analysis.
The technical problem of our approach stands in the difficulty of solving the equations of motion, which are, first of all, {\em operator-valued} and, secondly, quite
often nonlinear. In this paper we construct a general procedure which significantly simplifies the treatment for those systems which can be described in terms of fermionic operators.  The proposed procedure allows to get an
analytic solution, both for quadratic and for more general hamiltonians.

\end{abstract}

\vspace{2cm}


\vfill


\newpage

\section{Introduction and motivations}

In a series of recent papers we have discussed how several classical (complex) systems can be analyzed by adopting operatorial techniques which
are borrowed from quantum mechanics. We refer to \cite{bagbook} for a recent monograph concerning this  approach, which was used, in particular, for
economical,  biological and social systems. For instance, we have used this method in the description of stock markets, of the dynamics of
populations, of love affairs, all considered as dynamical systems. The dynamics is deduced by a self-adjoint operator, the hamiltonian $H$ of
the system $\Sc$, which describes the main effects which are observed in $\Sc$, \cite{bagbook,bagijtp}. Then, using an Heisenberg-like
procedure, the equations of motion are deduced: $\dot X(t)=i[H,X(t)]$, for each dynamical variable $X(t)$ whose dynamics is interesting for us.

In all the systems considered so far, the variables we were interested in (the so-called {\em observables} of $\Sc$) are number-like
operators, i.e. operators like $\hat N=a^\dagger\,a$, with $a$ and $a^\dagger$ respectively lowering and raising operators. In most cases,
these operators were assumed to satisfy the canonical commutation relations (CCR), $[a,a^\dagger]=\1$, see \cite{bag1,ff2} for example.
However, when discussing migration or, more generally, dynamics of populations, \cite{ff3}, it turns out to be more convenient to assume that $a$ satisfies
the canonical anti-commutation relations (CAR), $\{a,a^\dagger\}=\1$, $a^2=0$. The reason for this, is that $\hat N$ can be interpreted as the {\em
density} of the population somehow {\em attached} to $a$ and $a^\dagger$. This choice is useful also for technical reasons: the Hilbert space where the
model is defined is necessarily finite-dimensional, which implies that all the operators are nothing but finite matrices. This is very different from what
happens using bosons. The reason is simple: the orthonormal basis of eigenvectors of $\hat N=a^\dagger\,a$ is made by infinite, numerable, vectors if $a$ and
$a^\dagger$ satisfy the CCR, while it is just 2-dimensional if they satisfy CAR\footnote{Notice that we could also decide to use CCR and to somehow {\em cutoff} the related Hilbert space to work with a finite-dimensional version of it, which should, nevertheless, be large enough to capture the dynamics of the system we are considering. This is exactly what we have done in \cite{ff2}.}. However, this is not enough to get easily solvable models. In
particular, whenever nonlinear equations of motion are deduced out of the (non quadratic) hamiltonian, they have been solved so far by
adopting convenient numerical techniques, \cite{ff2}-\cite{ffam}. This imposes, of course, strong limitations on the kind of systems which can be efficiently analyzed using our strategy, otherwise also numerical schemes could not be sufficiently.

This problem is, at least in part, solved in this paper where we introduce a new, purely matricial, approach which allows a reasonably simple
treatment of those classical systems which can be described using CAR, independently of the expression of the hamiltonian $H$, and in
particular independently of the fact that $H$ is quadratic, cubic, quartic, and so on. This method is, as we will see, simple and easily
implemented using some mathematical software. It should be stressed already at this stage that what we produce in this way will not be a numerical solution of the dynamics of
the system, but the exact, analytical, solution corresponding to a certain set of initial conditions. Also, it is crucial to stress that with this approach we will be able to deal with systems whose dynamics is driven by nonlinear differential equations.

\vspace{2mm}

The paper is organized as follows: in the next section we describe the general strategy. In Section III we discuss two simple models, with
quadratic hamiltonian and linear, exactly solvable, differential equations. In this way we compare two solutions, one deduced by solving the differential equations and the other deduced by adopting our new idea. In Section IV we show that the same procedure works perfectly even
when $H$ is not quadratic. In particular, we discuss a model whose differential equations of motion can be solved easily, and we compare the
solution with what we get using our procedure. They are exactly the same. Then we consider two other nonlinear models whose differential
equations can only be solved numerically, and we show that, nevertheless, our procedure still works perfectly. Our conclusions are given in Section V.

\section{The technique}

Let $\Sc$ be a certain physical system whose dynamics we want to deduce, and let us suppose that it can be described in terms of $N$ different modes of fermionic
operators $a_j$, $j=1,2,\ldots,N$. This means that the CAR are satisfied: $\{a_j,a_k^\dagger\}=\delta_{j,k}\1$, together with $a_j^2=0$,
$j,k=1,2,\ldots,N$. As it is widely discussed in the literature, see \cite{rom} for instance, these operators can be represented as matrices acting on a $2^N$ dimensional Hilbert
space, $\Hil_N$: hence they are $2^N\times 2^N$ matrices. A natural orthonormal (o.n.) basis of $\Hil_N$,
$\F_N=\{\varphi_{i_{N-1},i_{N-2},\ldots,i_1,i_0},\,i_j=0,1, \, j=0,1,2\ldots,N-1\}$, is clearly made by $2^N$ vectors with $2^N$ components
each, eigenstates of the various number operators $\hat N_j=a_j^\dagger a_j$: $\hat N_j\varphi_{i_{N-1},i_{N-2},\ldots,i_1,i_0}=i_j\,\varphi_{i_{N-1},i_{N-2},\ldots,i_1,i_0}$. Let further $e_j$ be the $j$-th vector of the canonical basis $\E$ of $\Hil_N$, i.e. the vector with all zero entries except the $j$-th component, which
is one. Then each $\varphi_{i_{N-1},i_{N-2},\ldots,i_1,i_0}$ can be identified with a vector $e_j$, the one with
$j=2^0i_0+2^1i_1+\cdots+2^{N-2}i_{N-2}+2^{N-1}i_{N-1}+1$.  For instance, $\varphi_{0,0,\ldots,0,0}\equiv e_1$, $\varphi_{0,0,\ldots,0,1}\equiv e_2$,
$\varphi_{0,0,\ldots,1,1}\equiv e_4$, and so on. Sometimes in the following, to simplify the notation, we will indicate these vectors with
$\varphi_{\bf i}$, where ${\bf i}=(i_{N-1},i_{N-2},\ldots,i_1,i_0)$.

For fixed $N$ we get an unique o.n. basis $\F_N$ of eigenstates of $\hat N_j$, in terms of which the matrices $a_j$ and the adjoint $a_j^\dagger$ can be explicitly deduced.
Then, we can deduce the matrix representation for the number operator $\hat N_j=a_j^\dagger a_j$, simply by multiplying the two. Finally, since the
self-adjoint hamiltonian $H$ of $\Sc$ is constructed out of these matrices, see Sections III and IV, $H$ will be a $2^N\times 2^N$ matrix such
that $H=\overline{H }^T$: the complex conjugate of the transpose of $H$ coincides with $H$. Hence, $H$ can be surely diagonalized and, calling $\epsilon_j$ its
eigenvalues, we know that the related eigenstates can be used to construct an invertible matrix, $U$, such that
$$
UHU^{-1}=diag\{\epsilon_1, \epsilon_2, \epsilon_3,\ldots, \epsilon_{2^N}\}=:H_d,
$$
where not all the  eigenvalues are necessarily different. A simple Taylor expansion shows that $e^{iHt}=U^{-1}e^{iH_dt}U$, where now
$$
e^{iH_dt}=diag\{e^{i\epsilon_1t},e^{i\epsilon_2t},e^{i\epsilon_3t},\ldots,e^{i\epsilon_{2^N}t}\}
$$
is a diagonal matrix. This simple fact is crucial, since what is really interesting for us is just the time evolution of the number operators:
 $\hat N_j(t)=e^{iHt}\hat N_je^{-iHt}$, $j=1,2,\ldots,N$. This is, in fact, the solution of the Heisenberg equation already introduced in Section
I: $\frac{d}{dt} \hat N_j(t)=i[H,\hat N_j(t)]$, at least when $H$ is not explicitly time-dependent.

\vspace{2mm}

{\bf Remark:--} these same arguments could be repeated, in principle, for systems involving bosons, rather than fermions. However, in this case,
many technical and substantial difficulties arise, mainly due to the fact that the Hilbert space, in this case, is infinite dimensional.

\vspace{2mm}

Now, using the previous results, we have
$$
\hat N_j(t)=e^{iHt}\hat N_je^{-iHt}=U^{-1}e^{iH_dt}Ua_j^\dagger a_jU^{-1}e^{-iH_dt}U.
$$
Of course, if $U^\dagger=U^{-1}$, defining $b_j(t):=a_jU^{-1}e^{-iH_dt}U$, we conclude that $\hat N_j(t)=b_j^\dagger(t)b_j(t)$. Notice that
these new operators do not satisfy the CAR, since we can easily check that $\{b_j(t),b_k^\dagger(t)\}=a_ja_k^\dagger+e^{iHt}a_k^\dagger
a_je^{-iHt}\neq\delta_{j,k}$, in general.

The next step goes like this: since all we need to compute in our treatment is the mean value of $\hat N_j(t)$ on vectors which are eigenstates of the {\em initial} number operators, $\hat N_j(0)=\hat N_j$,
with eigenvalues corresponding to the initial conditions, see \cite{bagbook}, we get \be n_j(t):=\left<\varphi_{\bf in},\hat N_j(t)\varphi_{\bf
in}\right>=\|b_j(t)\varphi_{\bf in}\|^2.\label{21}\en Here $\varphi_{\bf in}$ is that particular vector of $\F_N$ corresponding to the initial
conditions. For example, if at $t=0$ the system (with $N=5$) has $n_1=n_3=0$, and all the other $n_j=1$, then $\varphi_{\bf
in}=\varphi_{1,1,0,1,0}$.

In the following two sections we will describe few fermionic models just from a mathematical point of view, to show how the above idea applies
but paying not much attention to the physical interpretation of these models. Only the last model, the {\em richest one},  see Section \ref{sectN=3nlconH0}, will also
be briefly considered in view of its applicative aspects.

\section{Examples with quadratic hamiltonians}

This section is dedicated to two simple models, one with $N=2$ and the other with $N=3$, for which all the computations can be carried out in
different ways, to show the equivalence of the strategies.

\subsection{Example 1: $N=2$}

We assume that the Hamiltonian $H$ of the system $\Sc$ can be written as $H=\lambda(a_2a_1^\dagger+a_1a_2^\dagger)$, where $\lambda$ is a real parameter, while $\{a_j,a_k^\dagger\}=\delta_{j,k}\1$, and $a_j^2=0$, $j=1,2$. A similar model, with $a_j$ satisfying CCR rather than CAR, was introduced in \cite{ff2} in the analysis of Love Affairs.

The differential equations of motion for the lowering operators, $\dot a_j(t)=i[H,a_j(t)]$, produce a very simple system, $$ \left\{
    \begin{array}{ll}
\dot a_1(t)=i\lambda a_2(t),\\
\dot a_2(t)=i\lambda a_1(t),\\
       \end{array}
        \right.$$
which can be solved analytically. Since $a_1(0)=a_1$ and $a_2(0)=a_2$, we find that $a_1(t)=a_1\cos(\lambda\,t)+ia_2\sin(\lambda\,t)$ and $a_2(t)=a_2\cos(\lambda\,t)+ia_1\sin(\lambda\,t)$. The initial status of $\Sc$ is described by a vector $\varphi_{n_1,n_2}=(a_1^\dagger)^{n_1} (a_2^\dagger)^{n_2}\varphi_{0,0}$. where $a_1\varphi_{0,0}=a_2\varphi_{0,0}=0$. Different choices of $n_j$ correspond to different initial conditions. Then
 \be \left\{
    \begin{array}{ll}
n_1(t)=\left<\varphi_{n_1,n_2},a_1^\dagger(t)a_1(t)\varphi_{n_1,n_2}\right>=n_1\cos^2(\lambda t)+n_2\sin^2(\lambda t),\\
n_2(t)=\left<\varphi_{n_1,n_2},a_2^\dagger(t)a_2(t)\varphi_{n_1,n_2}\right>=n_2\cos^2(\lambda t)+n_1\sin^2(\lambda t).\\
       \end{array}
        \right.\label{31}\en
Incidentally we see that $n_1(t)+n_2(t)=n_1+n_2$: the sum of the densities of the two species is preserved during the time evolution. This suggests, see \cite{bagbook}, that an operator exists which commutes with $H$. In fact, we can check that $[H,\hat N_1+\hat N_2]=0$.

\subsubsection{Our look to this same model}\label{sect31bis}

What we have done analytically, solving a (simple) system of coupled differential equations, we want to do now using the general ideas introduced in Section II, and we want to compare the results.

The first step consists in deducing the matrix expression for $H$. For that we use the following 4-dimensional representation of the CAR algebra:
$$
a_1=\1_2\otimes\sigma_+=\left(
      \begin{array}{cccc}
        0 & 1 & 0 & 0 \\
        0 & 0 & 0 & 0 \\
        0 & 0 & 0 & 1 \\
        0 & 0 & 0 & 0 \\
      \end{array}
    \right),\qquad a_2=\sigma^+\otimes\sigma_z=\left(
      \begin{array}{cccc}
        0 & 0 & 1 & 0 \\
        0 & 0 & 0 & -1 \\
        0 & 0 & 0 & 0 \\
        0 & 0 & 0 & 0 \\
      \end{array}
    \right),$$ where $\1_2$ is the $2\times2$-identity matrix, while $\sigma_+=\left(
                                                                                 \begin{array}{cc}
                                                                                   0 & 1 \\
                                                                                   0 & 0 \\
                                                                                 \end{array}
                                                                               \right)
    $ and $\sigma_z=\left(
                                                                                 \begin{array}{cc}
                                                                                   1 & 0 \\
                                                                                   0 & -1 \\
                                                                                 \end{array}
                                                                               \right)$ are two Pauli matrices. The vectors of $\F_2$ are $$\varphi_{0,0}=\left(
                                \begin{array}{c}
                                  1 \\
                                  0 \\
                                  0 \\
                                  0 \\
                                \end{array}
                              \right),\varphi_{0,1}=\left(
                                \begin{array}{c}
                                  0 \\
                                  1 \\
                                  0 \\
                                  0 \\
                                \end{array}
                              \right),\quad \varphi_{1,0}=\left(
                                \begin{array}{c}
                                  0 \\
                                  0 \\
                                  1 \\
                                  0 \\
                                \end{array}
                              \right),\quad
                              \varphi_{1,1}=\left(
                                \begin{array}{c}
                                  0 \\
                                  0 \\
                                  0 \\
                                  1 \\
                                \end{array}
                              \right).
$$
They are mutually orthogonal and normalized. Also, they satisfy the standard relations: $a_j\varphi_{0,0}=0$, $j=1,2$, $a_1^\dagger\varphi_{0,0}=\varphi_{0,1}$, $a_2^\dagger\varphi_{0,0}=\varphi_{1,0}$, and $a_1^\dagger a_2^\dagger\varphi_{0,0}=\varphi_{1,1}$. $H$ is now represented by the following symmetric (and self-adjoint) matrix:
$$
H=\lambda\left(
      \begin{array}{cccc}
        0 & 0 & 0 & 0 \\
        0 & 0 & -1 & 0 \\
        0 & -1 & 0 & 0 \\
        0 & 0 & 0 & 0 \\
      \end{array}
    \right)
$$
which can be easily diagonalized: the four eigenvalues are $\epsilon_1=\epsilon_2=0$, $\epsilon_3=-\lambda$, $\epsilon_4=\lambda$.
The related eigenvectors are $\eta_1^T=(0,0,0,1)$, $\eta_2^T=(1,0,0,0)$, $\eta_3^T=\frac{1}{\sqrt{2}}(0,0,1,-1)$,
 $\eta_4^T=\frac{1}{\sqrt{2}}(0,0,1,1)$, so that $$U^{-1}=U^\dagger=\left(
      \begin{array}{cccc}
        0 & 1 & 0 & 0 \\
        0 & 0 & 1/\sqrt{2} & -1/\sqrt{2} \\
        0 & 0 & 1/\sqrt{2} & 1/\sqrt{2} \\
        1 & 0 & 0 & 0 \\
      \end{array}
    \right).
$$
Now, recalling that $b_j(t):=a_jU^{-1}e^{-iH_dt}U$, we find
$$
b_1(t)=\left(
      \begin{array}{cccc}
        0 & \cos(\lambda t) & i\sin(\lambda t) & 0 \\
        0 & 0 & 0 & 0 \\
        0 & 0 & 0 & 1 \\
        0 & 0 & 0 & 0 \\
      \end{array}
    \right),\qquad b_2(t)=\left(
      \begin{array}{cccc}
        0 & i\sin(\lambda t) & \cos(\lambda t) & 0 \\
        0 & 0 & 0 & -1 \\
        0 & 0 & 0 & 0 \\
        0 & 0 & 0 & 0 \\
      \end{array}
    \right).
$$
Using equation (\ref{21}), we can finally find the expressions of $n_j(t)$ corresponding to different initial conditions. This is nothing than a computation of the norm of some vectors: for instance, if at $t=0$ the system $\Sc$ has $n_1=1$ and $n_2=0$, then $n_1(t)=\|b_1(t)\varphi_{0,1}\|^2=\cos^2(\lambda t)$, while $n_2(t)=\|b_2(t)\varphi_{0,1}\|^2=\sin^2(\lambda t)$. This same result can be deduced using (\ref{31}).

\subsection{Example 2: $N=3$}\label{sect32bis}

This example extends the previous one, meaning with this that it is based on the existence of 3, and not just 2, different fermionic modes. The hamiltonian is
$$H=\lambda(a_2a_1^\dagger+a_1a_2^\dagger+a_3a_1^\dagger+a_1a_3^\dagger+a_3a_2^\dagger+a_2a_3^\dagger),$$ where, again, $\lambda$ is a real
parameter, and $\{a_j,a_k^\dagger\}=\delta_{j,k}\1$, and $a_j^2=0$, $j,k=1,2,3$.

The Heisenberg equations of motion are
 $$ \left\{
    \begin{array}{ll}
\dot a_1(t)=i\lambda (a_2(t)+a_3(t)),\\
\dot a_2(t)=i\lambda (a_1(t)+a_3(t)),\\
\dot a_3(t)=i\lambda (a_1(t)+a_2(t)),\\
       \end{array}
        \right.$$
which are linear and can be solved analytically. After some manipulations, we deduce the following:
\be
\left\{
    \begin{array}{ll}
n_1(t)=\left<\varphi_{n_1,n_2},a_1^\dagger(t)a_1(t)\varphi_{n_1,n_2}\right>=n_1|e_{1,1}(t)|^2+n_2|e_{1,2}(t)|^2+n_3|e_{1,3}(t)|^2,\\
n_2(t)=\left<\varphi_{n_1,n_2},a_2^\dagger(t)a_2(t)\varphi_{n_1,n_2}\right>=n_1|e_{2,1}(t)|^2+n_2|e_{2,2}(t)|^2+n_3|e_{2,3}(t)|^2,\\
n_3(t)=\left<\varphi_{n_1,n_2},a_3^\dagger(t)a_3(t)\varphi_{n_1,n_2}\right>=n_1|e_{3,1}(t)|^2+n_2|e_{3,2}(t)|^2+n_3|e_{3,3}(t)|^2,\\
       \end{array}
        \right.\label{32}\en
where
$$
\left\{
    \begin{array}{ll}
e_{j,j}(t)=\frac{1}{3}\left[2\cos(\lambda t)+\cos(2\lambda t)+i(-2\sin(\lambda t)+\sin(2\lambda t))\right)],\\
e_{j,k}(t)=\frac{1}{3}\left[-\cos(\lambda t)+\cos(2\lambda t)+i(\sin(\lambda t)+\sin(2\lambda t))\right], \quad j\neq k,\\
       \end{array},
        \right.
$$
$j,k=1,2,3$. Again, an integral of motion exists, and this is just the {\em global} number operator $\hat N=\hat N_1+\hat N_2+\hat N_3$: $[H,\hat N]=0$: whenever a fermion of mode 1 is destroyed, another (in mode 2 or 3) must be created, and viceversa.

\subsubsection{Our look to this same model}

As in the previous example, the first step consists in deducing the matrix expression for $H$. For that we use the following representation of the CAR algebra:
$$
a_1=\1_2\otimes\1_2\otimes\sigma_+,\qquad a_2=\1_2\otimes\sigma^+\otimes\sigma_z, \qquad a_3=\sigma^+\otimes\sigma_z\otimes\sigma_z,$$
which are $8\times8$ matrices. The o.n. basis $\F_3=\{\varphi_{i_2,i_1,i_0},\,i_j=0,1,\, j=0,1,2\}$, which extends that of the previous example, is the canonical basis in ${\Bbb C}^8$. In this basis the hamiltonian is
$$
H=-\lambda\left(
            \begin{array}{cccccccc}
              0 & 0 & 0 & 0 & 0 & 0 & 0 & 0 \\
              0 & 0 & 1 & 0 & 1 & 0 & 0 & 0 \\
              0 & 1 & 0 & 0 & 1 & 0 & 0 & 0 \\
              0 & 0 & 0 & 0 & 0 & 1 & -1 & 0 \\
              0 & 1 & 1 & 0 & 0 & 0 & 0 & 0 \\
              0 & 0 & 0 & 1 & 0 & 0 & 1 & 0 \\
              0 & 0 & 0 & -1 & 0 & 1 & 0 & 0 \\
              0 & 0 & 0 & 0 & 0 & 0 & 0 & 0 \\
            \end{array}
          \right)
$$
whose eigenvalues are $\epsilon_1=\epsilon_2=0$,  $\epsilon_3=-2\lambda$, $\epsilon_4=\epsilon_5=-\lambda$, $\epsilon_6=\epsilon_7=\lambda$,
$\epsilon_8=2\lambda$. Using as before the orthonormal eigenvectors of $H$ we can construct the matrix $U^{-1}$, and $U$ as a consequence. We
get
$$
U^{-1}=\left(
            \begin{array}{cccccccc}
              0 & 1 & 0 & 0 & 0 & 0 & 0 & 0 \\
              0 & 0 & \sqrt{\frac{1}{3}} & 0 & 0 & -\sqrt{\frac{1}{2}} & -\sqrt{\frac{1}{6}} & 0 \\
              0 & 0 & \sqrt{\frac{1}{3}} & 0 & 0 & 0 & \sqrt{\frac{2}{3}} & 0 \\
              0 & 0 & 0 & -\sqrt{\frac{1}{2}} & \sqrt{\frac{1}{6}} & 0 & 0 & \sqrt{\frac{1}{3}} \\
              0 & 0 & \sqrt{\frac{1}{3}} & 0 & 0 & \sqrt{\frac{1}{2}} & -\sqrt{\frac{1}{6}} & 0 \\
              0 & 0 & 0 & 0 & \sqrt{\frac{2}{3}} & 0 & 0 & -\sqrt{\frac{1}{3}} \\
              0 & 0 & 0 & \sqrt{\frac{1}{2}} & \sqrt{\frac{1}{6}} & 0 & 0 & \sqrt{\frac{1}{3}} \\
              1 & 0 & 0 & 0 & 0 & 0 & 0 & 0 \\
            \end{array}
          \right),
$$
while $b_j(t)$ are deduced as in Section II. For instance we get
$$
b_1(t)=\left(
            \begin{array}{cccccccc}
              0 & \frac{2e^{-i t \lambda }+e^{2 i t \lambda }}{3}  & \frac{-e^{-i t \lambda }+e^{2 i t \lambda }}{3}   & 0 & \frac{-e^{-i t \lambda }+e^{2 i t \lambda }}{3}   & 0 & 0 & 0 \\
              0 & 0 & 0 & 0 & 0 & 0 & 0 & 0 \\
              0 & 0 & 0 & \frac{e^{-2 i t \lambda }+2 e^{ i t \lambda }}{3}   & 0 & \frac{-e^{-2 i t \lambda }+e^{ i t \lambda }}{3}   & -\frac{-e^{-2 i t \lambda }+e^{ i t \lambda }}{3}   & 0 \\
              0 & 0 & 0 & 0 & 0 & 0 & 0 & 0 \\
              0 & 0 & 0 & \frac{-e^{-2 i t \lambda }+ e^{ i t \lambda }}{3} & 0 & \frac{e^{-2 i t \lambda }+2e^{ i t \lambda }}{3} & \frac{-e^{-2 i t \lambda }+e^{ i t \lambda }}{3} & 0 \\
              0 & 0 & 0 & 0 & 0 & 0 & 0 & 0 \\
              0 & 0 & 0 & 0 & 0 & 0 & 0 & 1 \\
              0 & 0 & 0 & 0 & 0 & 0 & 0 & 0 \\
            \end{array}
          \right),
$$
and similar expressions can be found for $b_2(t)$ and $b_3(t)$. We are now ready to compute $n_j(t)$ for different initial conditions. For
instance, if $n_1=n_2=n_3=0$, then $n_j(t)=\|b_j(t)\varphi_{0,0,0}\|^2=0$, $j=1,2,3$. Analogously, if $n_1=1$, $n_2=n_3=0$, then
$n_j(t)=\|b_j(t)\varphi_{0,0,1}\|^2$ and we get, for instance, $n_1(t)=\frac{1}{9}(5+4\cos(3\lambda t))$. Also, if $n_1=n_2=1$ and $n_3=0$ we find
$n_1(t)=n_2(t)=\frac{1}{9}(7+2\cos(3\lambda t))$, while $n_3(t)=\frac{8}{9}\sin^2\left(\frac{3\lambda t}{2}\right)$.

It is a simple exercise to check that these results (as well as the others corresponding to different initial conditions) coincide with those
in (\ref{32}).

\section{Examples with cubic hamiltonians}

The examples considered in the previous section are useful mainly because they suggest that what we are doing here is equivalent to
what we have done in our previous applications. Generalizing what we have deduced so far, we could claim that, as far as the differential
equations of motion are linear, the two approaches, let's call them {\em differential} and {\em matricial}, are equivalent. These examples  also show that, when it is possible, it is much more convenient to use the differential rather than the matricial approach, for instance because a single formula contains all the possible results for all possible different initial values. However, the differential approach cannot be always carried out. With this in mind, in this section we
will discuss what happens when the differential equations are no longer linear. In particular, we will consider first a model for which the
differential approach can again be considered, and we check that the solution we get coincides with that obtained adopting the matricial technique.
After that, we consider two models for which the differential equations can only be solved numerically, showing that our matricial technique
still applies and produces an explicit solution.

\subsection{N=2: a solvable model}

Let us consider the  hamiltonian: $H=\lambda(a_1^\dagger \hat N_2+\hat N_2 a_1)$, where $a_j$, $a_j^\dagger$ are the usual fermionic operators
and $\hat N_2=a_2^\dagger a_2$. Since $[H,\hat N_2]=0$, it follows that $\hat N_2(t)=\hat N_2(0)=\hat N_2$ and $n_2(t)=\left<\varphi_{n_1,n_2},\hat
N_2(t)\varphi_{n_1,n_2}\right>=n_2$. On the other hand, since $[H,\hat N_1]\neq0$, $\hat N_1(t)\neq \hat N_1(0)$.  In order to deduce $\hat N_1(t)$, and its mean value $\left<\varphi_{n_1,n_2},\hat
N_1(t)\varphi_{n_1,n_2}\right>$, it is convenient to look for the differential equation for $a_1(t)$: $\dot
a_1(t)=i\lambda (2\hat N_1(t)-\1)\hat N_2$. Its adjoint is $\dot
a_1^\dagger(t)=-i\lambda (2\hat N_1(t)-\1)\hat N_2$. Then $a_1(t)+a_1^\dagger(t)=a_1+a_1^\dagger$, for all $t\geq0$. Moreover, since
$\frac{d}{dt}\hat N_1(t)=\dot a_1^\dagger(t)a_1(t)+a_1^\dagger(t)\dot a_1(t)$, we deduce that $ \frac{d}{dt}\hat N_1(t)=i\lambda \hat
N_2(a_1(t)-a_1^\dagger(t))$, and therefore $$\frac{d^2}{dt^2}\hat N_1(t)+4\lambda^2\hat N_2\hat N_1(t)=2\lambda^2 \hat N_2.$$ A simple analysis
of this equation produces the following solutions, depending on the initial conditions: (i) if $n_1=n_2=0$ then $n_1(t)=n_2(t)=0$; (ii) if
$n_1=1$ and $n_2=0$ then $n_1(t)=1$ and $n_2(t)=0$; (iii) if $n_1=0$ and $n_2=1$ then $n_1(t)=\sin^2(\lambda t)$ and $n_2(t)=1$; (iv) if
$n_1=1$ and $n_2=1$ then $n_1(t)=\cos^2(\lambda t)$ and $n_2(t)=1$, for all $t\geq0$.

\subsubsection{Our look to this same model}

As before, we look for the matrix expression for $H$. For that we use the matrix expressions for $a_1$ and $a_2$ introduced in Section \ref{sect31bis} and
we find that
$$
H=\lambda\left(
      \begin{array}{cccc}
        0 & 0 & 0 & 0 \\
        0 & 0 & 0 & 0 \\
        0 & 0 & 0 & 1 \\
        0 & 0 & 1 & 0 \\
      \end{array}
    \right).
$$
This matrix can be diagonalized quite easily: the eigenvalues are $0,0,\pm \lambda$, and the matrix $U^{-1}$ has the following form:
$$U^{-1}=U^\dagger=\left(
      \begin{array}{cccc}
        0 & 1 & 0 & 0 \\
        1 & 0 & 0 & 0 \\
        0 & 0 & -1/\sqrt{2} & 1/\sqrt{2} \\
        0 & 0 & 1/\sqrt{2} & 1/\sqrt{2} \\
      \end{array}
    \right).
$$
Recalling that $b_j(t):=a_jU^{-1}e^{-iH_dt}U$, we find
$$
b_1(t)=\left(
      \begin{array}{cccc}
        0 & 1 & 0  & 0 \\
        0 & 0 & 0 & 0 \\
        0 & 0 & -i\sin(\lambda t) & \cos(\lambda t) \\
        0 & 0 & 0 & 0 \\
      \end{array}
    \right),\qquad b_2(t)=\left(
      \begin{array}{cccc}
        0 & 0 & \cos(\lambda t) & -i\sin(\lambda t)0 \\
        0 & 0 & i\sin(\lambda t) & \cos(\lambda t) \\
        0 & 0 & 0 & 0 \\
        0 & 0 & 0 & 0 \\
      \end{array}
    \right).
$$
Now, using (\ref{21}), we could find the expressions of $n_j(t)$ corresponding to different initial conditions.  Again, this is nothing than a
computation of the norm of some vectors. The results coincide, as expected, with those which were already deduced.

\subsection{N=3: a model with no $H_0$}

The example we are going to consider now is different, with respect to those we have considered so far, since the differential approach does
not apparently produce any analytical solution. Only a numerical scheme, or some perturbation expansion, can be used to solve the differential equations in (\ref{41})
below.

The hamiltonian is $H=\lambda(a_1^\dagger a_2^\dagger a_3+a_3^\dagger a_2 a_1)$, where $\lambda$ is, as usual, a real parameter and $a_j$ are
fermionic operators satisfying CAR. The Heisenberg equations of motion are \be \left\{
    \begin{array}{ll}
\dot a_1(t)=i\lambda [a_1^\dagger(t),a_1(t)]a_2^\dagger(t)a_3(t),\\
\dot a_2(t)=-i\lambda a_1^\dagger(t)[a_2^\dagger(t),a_2(t)]a_3(t),\\
\dot a_3(t)=-i\lambda a_1(t)a_2(t) [a_3^\dagger(t),a_3(t)],\\
       \end{array}
        \right.\label{41}\en
and an analytic solution seems not to be easily found. On the other hand, our idea trivially applies. In fact, adopting the representation used in Section
\ref{sect32bis}, the hamiltonian can be written as
$$
H=\lambda\left(
\begin{array}{cccccccc}
 0 & 0 & 0 & 0 & 0 & 0 & 0 & 0 \\
 0 & 0 & 0 & 0 & 0 & 0 & 0 & 0 \\
 0 & 0 & 0 & 0 & 0 & 0 & 0 & 0 \\
 0 & 0 & 0 & 0 & 1  & 0 & 0 & 0 \\
 0 & 0 & 0 & 1  & 0 & 0 & 0 & 0 \\
 0 & 0 & 0 & 0 & 0 & 0 & 0 & 0 \\
 0 & 0 & 0 & 0 & 0 & 0 & 0 & 0 \\
 0 & 0 & 0 & 0 & 0 & 0 & 0 & 0
\end{array}
\right).
$$
The operators $b_j(t)$ assume simple expressions. For instance we get
$$
b_1(t)=\left(
            \begin{array}{cccccccc}
              0 & 1 & 0 & 0 & 0  & 0 & 0 & 0 \\
              0 & 0 & 0 & 0 & 0 & 0 & 0 & 0 \\
              0 & 0 & 0 & \cos(\lambda t)  & -i\sin(\lambda t) & 0  & 0  & 0 \\
              0 & 0 & 0 & 0 & 0 & 0 & 0 & 0 \\
              0 & 0 & 0 & 0 & 0 & 1 & 0 & 0 \\
              0 & 0 & 0 & 0 & 0 & 0 & 0 & 0 \\
              0 & 0 & 0 & 0 & 0 & 0 & 0 & 1 \\
              0 & 0 & 0 & 0 & 0 & 0 & 0 & 0 \\
            \end{array}
          \right),
$$
and so on. The time evolution of the mean values of the number operators is quite easily found using (\ref{21}), and of course depends on the
initial conditions: (i) if $n_1=n_2=n_3=0$, then $n_j(t)=0$; (ii) if $n_1=1$ and $n_2=n_3=0$, then $n_1(t)=1$, $n_2(t)=n_3(t)=0$; (iii) if
$n_2=1$ and $n_1=n_3=0$, then $n_2(t)=1$, $n_1(t)=n_3(t)=0$; (iv) if $n_3=1$ and $n_1=n_2=0$, then $n_1(t)=n_2(t)=\sin^2(\lambda t)$,
$n_3(t)=\cos^2(\lambda t)$; (v) if $n_3=0$ and $n_1=n_2=1$, then $n_1(t)=n_2(t)=\cos^2(\lambda t)$, $n_3(t)=\sin^2(\lambda t)$, and so on.
These results show that $n_{13}(t)=n_1(t)+n_3(t)$ and $n_{23}(t)=n_2(t)+n_3(t)$ stay constant in time. This is in agreement with the fact that
both $\hat N_1+\hat N_3$ and $\hat N_2+\hat N_3$ commute with $H$. The {\em physical} reason for this is easily understood, looking at the
explicit expression for $H$: in fact, $H$ contains the contribution $a_1^\dagger a_2^\dagger a_3$, which implies that, whenever a fermion of type 3 is
annihilated, one of the type 1 and another one of type 2 are created. The adjoint term in $H$ describes a specular phenomenon (a type-3 fermion is created and two type-1 and type-2 fermions are annihilated), which again
preserves $n_{13}$ and $n_{23}$, but not the total number of fermions.

\vspace{3mm}

This particular example shows that, an apparently very difficult problem, as the one represented by the equations in (\ref{41}), can be
efficiently treated working in matrix terms. The same conclusion will be deduced in the next example, where a {\em standard} free part,
\cite{bagbook}, will be added to $H$.

\subsection{N=3: a model with $H_0$}\label{sectN=3nlconH0}

The hamiltonian we consider now is the following extension of the previous one:
$$
H=\omega(a_1^\dagger a_1+a_2^\dagger a_2+a_3^\dagger a_3)+\lambda(a_1^\dagger a_2^\dagger a_3+a_3^\dagger a_2 a_1).
$$
The reason to use a single free parameter $\omega>0$ is to make the situation simpler. In
general, however, there is a different $\omega$ for each fermionic mode. The interpretation
of these parameters in realistic models is discussed in \cite{bagbook}. The hamiltonian above could be used to describe two biological species (modes 1 and
2) and the food used to feed them (mode 3): the more the species grow, the larger is the amount of food to be used. This is the meaning of
$a_1^\dagger a_2^\dagger a_3$. On the contrary, when the densities of the species decrease, the food is not used so much, and, therefore, it
can increase. This is why $a_3^\dagger a_2 a_1$ appears in $H$. The free term, $\omega(a_1^\dagger a_1+a_2^\dagger a_2+a_3^\dagger a_3)$, when
there is no interaction ($\lambda=0$), describes a stationary situation which keeps all the densities constant in time, \cite{bagbook}.

\vspace{2mm}

The Heisenberg equations of motion extend those in (\ref{41}), making them even more complicated. Not surprisingly, therefore, we are not able
to solve them analytically, even considering the fact that, as when $\omega=0$, $[H,\hat N_1+\hat N_3]= [H,\hat N_2+\hat N_3]=0$.

On the other hand, we can still adopt our simple strategy. For instance, fixing $\omega=1$ and $\lambda=0.1$, we find quite easily that, for instance, if $n_3=1$ and $n_1=n_2=0$, then
$n_1(t)=n_2(t)=0.0192308-0.0192308 \cos(1.0198 t)$, $n_3(t)=0.980769+0.0192308 \cos(1.0198 t)$. Similar results can be found for different
initial condition and for different choices of the parameters. The conclusion is the same as before: even when the differential equations cannot be analytically solved, as quite often is the case for non purely quadratic hamiltonians, our strategy still produces the solution. This is quite interesting mainly in view of future, more realistic, applications. Also, the results are deduced in a very small amount of time, and they are not very depending on the nonlinearity of the differential equations which come
out from the Hamiltonian of the system.

\section{Final remarks and conclusions}

In this short note we have introduced and adopted a simple method to deduce the dynamics of some system described in terms of fermionic
operators. We have seen that our method works quite well independently of the nature of the hamiltonian $H$ of the system, $\Sc$. The examples
presented here are reasonably simple and not particularly interesting for concrete applications. Our next step will consist in using our
technique in more interesting models, as those already discussed in \cite{ff3} and in \cite{fff,ffam}, for which the hamiltonians appear to be significantly more complicated.

We end the paper with a no-go result which, nevertheless, opens possible lines of research for the future: what we have discussed here works when $\Sc$ is
described in terms of fermionic operators. But it cannot work when $\Sc$ {\em needs bosons}. In this case, we still have to find a way
to simplify the analysis. Of course, an approximated procedure is easily implemented: if we cutoff the infinite-dimensional Hilbert space
$\Hil$, by considering an effective space $\Hil_{eff}$, then the observables of $\Sc$ are replaced by matrices on $\Hil_{eff}$, and the same
technique described in Section II can be, in principle, adopted. This cutoff procedure was used successfully in \cite{ff2}, and was analytically justified because of the
existence of a certain integral of motion. We believe that, when such an integral exists, a similar approximation can again be implemented, and
therefore the solution can be deduced as in Section II. A deeper analysis on these aspects is in progress.

\section*{Acknowledgements}

This work was partially supported by the University of Palermo.


\end{document}